\documentclass{elsart}

\usepackage{graphicx}

\usepackage{amssymb}

\begin{document}

\begin{frontmatter}



\title{Variability of Broad and Blueshifted Component of [OIII]$\lambda$5007 
  in I\,ZWI} 


\author{J. Wang}
\address{Department of Astronomy, Beijing Normal University, 
          NO.19, Xinjiekouwai St, Beijing, China}
\ead{wj@bao.ac.cn}

\author{J. Y. Wei}
\address{National Astronomical Observatories, Chinese Academy of Sciences,
          20A Datun Road, Chaoyang District, Beijing, China}
\ead{wjy@bao.ac.cn}

\author{X. T. He}
\address{Department of Astronomy, Beijing Normal University,
         NO.19, Xinjiekouwai St, Beijing, China}
\ead{xthe@bnu.edu.cn}

\begin{abstract}
   Although the existence of asymmetrical profile of [OIII]$\lambda$5007
   has been discovered for ages, its filiation and physics are poorly understood.
   Two new spectra of I\,ZWI taken on Nov 16, 2001 and on Dec 3, 2002 were compared 
   with the spectra taken by BG92. Following results are obtained. 1)The certain 
   variations of broad [OIII] during about 10 years separating the observations are 
   identified. The inferred length scale of broad [OIII] emitting region 
   ranges from 0.3pc to 3pc. By assuming a Keplerian motion in emitting region,
   the material emitting broad [OIII] is likely to be located at transient emission 
   line region, between BLR and NLR. 2)We find a positive relation between the 
   FeII emission and flux of H$\beta$(or continuum). On the other hand, 
   the parameter RFe decreases with ionizing continuum marginally. 3)We detect a 
   low ionized NLR in I\,ZWI, because of the low flux 
   ratios $\rm{[OIII]_{n}/H\beta_{n}}$($\sim1.7$). 

\end{abstract}

\end{frontmatter}

\section{Introduction}
   The asymmetry of [OIII]$\lambda$5007 profile possessing an extended wing at blueward
   and a sharp red falloff was mentioned by Heckman et al.(1981) and by 
   many subsequent researches(e.g. Busko \& Steiner 1988, Grupe et al. 1999,  Holt et al. 2003, 
   Tadhunter et al. 2001). On the basis of investigations on large sample, recent studies claimed that 
   a large fraction of AGNs are of [OIII] broad components. V\'{e}ron-Cetty et al.(2001) reported that,
   in general, a relative narrow Gaussian profile(FWHM$\sim \rm{200-500\ km\ s^{-1}}$) with a blueshifted,
   broad Gaussian component(FWHM$\sim \rm{500-1800\ km\ s^{-1}}$) is necessary to reproduce the 
   each [OIII] profile in a number of objects(24). 
   In addition, a peak + blue wing structure was observed in numerous objects by Zamanov et al.(2002).
   The authors claimed that the [OIII] profiles have a narrow, unshifted [OIII] line with a strong blue 
   wing. Recently, Wang et al.(2004) identified a prominent blueshifted, broad [OIII] component in 
   Narrow-Line Seyfert 1 galaxy(NLS1) SDSS\,J022119.84+005628.4.

   So far, although the peak+blue wing structure of [OIII] has been detected 
   in many cases, the properties of the broad [OIII] component have been rarely investigated and 
   poorly understood. To our knowledge, only Sergeev et al.(1997) analyzed the variations of [OIII]
   $\lambda$5007 blue wing in NGC\,5548. The transfer function of the blue wing is narrow-peaked near 450 days
   with no response up to 350 days lag. One interesting question is where is the broad component of [OIII]$\lambda5007$
   from and how large is the length scale of the line emitting region.
   The answers are extremely benefit to further geometrical and dynamical research of
   emission line region in AGN.

   As usual, the study of spectral variations is a powerful technology in  
   probing the physics of AGN. Variability can help us to set constraints on the 
   sizes of different regions of AGNs and can give information about the 
   processes governing the variations. The object, I\,ZWI, is the prototype of
   NLS1 and was discovered by Zwicky(1971). In this paper we present two new spectra of 
   I\,ZWI taken in 2001 and 2002. The variations of the [OIII] broad component are
   examined by comparing them with the spectrum taken by Boroson \& Green(1992, hereafter BG92).
   They examined the emission line properties of a sample of 87 low redshift ($z<0.5$) PG quasars.
   The spectra with relatively high S/N ratio were obtained on a number of nights in 1990 and 1991.
   A resolution of 6.5-7{\AA} was measured from the comparison spectra.
   The resolution at this level is adequate for identifying the evident asymmetry of the [OIII]$\lambda5007$ line.

   The paper is organized as follows. The observations and date reduction technique are described in \S2.
   Section 3 contains the results of spectral comparison and underlying implications.

\label{Intro}

\section{Observation and data reduction}

   The two spectra were taken on November 16, 2001 and December 3, 2002. The observations 
   were carried out with the NAOC 2.16m telescope at the observatory of Xinglong and the 
   OMR spectrograph, using a Tektronix $1024\times1024$ CCD as detector.
   Two different 600$\rm{l\ mm^{-1}}$ gratings, blazed at 5500{\AA} and 5000{\AA},
   were used in observation taken in 2001 and in 2002 respectively. Both grating give a dispersion 
   of about 100$\AA\ \rm{mm^{-1}}$. The slit was oriented in the north-south direction for both 
   observations and we attempted to observe as close to the meridian as possible.
   Both observations were made through a 2" slit which produced 
   a resolution of $\sim$ 6\AA\ as measured from the night sky lines. This set-up caused that 
   the spectral resolution is identical with that of BG92 at greatest degree. 
   In the observation taken in 2001, two exposures were performed. The exposure time of 
   each frame was 2000s. The two frames were combined prior to extraction in order to
   enhance the S/N ratio and eliminate the contamination of cosmic-ray easily. 
   Only one frame with exposure time 3000s was obtained in the observation taken in 2002.

   The unprocessed frames were reduced by standard CCD procedure using IRAF package. The CCD reductions
   include bias subtraction, flatfield correction and cosmic-ray removal. The wavelength
   calibration was carried out using helium-neon-argon lamps taken at the beginning and end of each exposure.
   The resulting wavelength accuracy is better than 1\AA.  
   Two or three KPNO standard
   stars(Massey et al. 1988) were observed per night for carrying out the flux calibration.
   Each of extracted spectra was transformed to rest frame in terms of the redshift determined by fitting 
   the narrow peak of the H$\beta$ line to a Gaussian.

   \subsection{FeII subtraction and measurements}

   It is clear from the inspection of the spectra that the FeII emission is a complicating factor 
   in measuring line parameters. In order to reliably model the line profiles it is necessary to 
   appropriately model the FeII complex. The contamination of the FeII complex is 
   well subtracted by the FeII template which is the FeII emission in I\,ZWI and is described by BG92.  
   Briefly, the template is a two-dimensional function of FWHM and of intensity of the FeII blends.
   The detailed procedure making the FeII template can be found in BG92. The best subtraction of 
   the FeII blends is derived by searching in the parameter space and requires a slick
   continuum at blue of H$\beta$ and between 5100{\AA} and 5500{\AA}. From each spectrum 
   the best-fit FeII spectrum is subtracted. These FeII spectra and the resulting FeII-subtracted 
   spectra are shown in Figure 1. The spectra displayed from top to bottom panel were observed by BG92,
   in Nov 16, 2001(hereafter denoted as WWHI for abbreviation) and Dec 03, 2002(WWHII, for short), 
   respectively. In each panel, the observed spectrum is displayed by the upper curve.
   Note that these curves are offset upward arbitrarily for visibility. The FeII flux is measured between 
   the rest wavelength 4434\AA\ and 4684\AA. The upper and lower limits of the flux of the FeII
   complex are obtained by carefully iterative experiments with a series of values of the FeII flux.
   Outside of the limits, the FeII-subtracted continuum is absolutely unacceptable.

   \subsection{Profile modelling and decomposition of other lines }

    The FeII-subtracted spectrum are used to measure the non-FeII line properties. 
    The first step in the modelling of the spectra is to remove the continuum from 
    each FeII-subtracted spectrum. The continuum is modelled by fitting a power law 
    to the regions which seemed to be uncontaminated by emission lines. In the next step,
    the line profiles are modelled by multiple Gaussian fitting in the present work(e.g. Xu et al. 2003).
    The modelling is carried out by the SPECFIT(Kriss 1994) task in IRAF package. 
    The fitting is persisted until the minimum of $\chi^2$, the measurement of goodness of the
    fitting, is achieved. In each spectrum, the profiles are modelled as follows.
    The each of forbidden lines [OIII]$\lambda\lambda$4959,5007 is synthesized from a narrow 
    component and a broad, blueshifted component(see Oke \& Lauer 1979, V\'{e}ron-Cetty et al. 2004).
    The atomic physical relationships, $F_{5007}/F_{4959}\doteq3$(Storey \& Zeippen 2000) and
    $\lambda_{4959}/\lambda_{5007}=0.9904$, are used for reducing the number of free parameters
    and for improving the reliability of fitting in the modelling of both narrow and broad components.
    The narrow component of [OIII] is referred as $\rm{[OIII]_{n}}$, and the broad component as $\rm{[OIII]_{b}}$ 
    The H$\beta$ profile is reproduced by a set of three components: a narrow peak, a classical broad 
    component with FWHM$\sim 1000\ \rm{km\ s^{-1}}$ and a very broad base (FWHM$\sim 4000\ \rm{km\ s^{-1}}$,
    see Sulentic et al. 2000b; V\'{e}ron-Cetty et al. 2001; Marziani et al. 2003), 
    because of the strong blue wing of the H$\beta$ profile. The width of the narrow H$\beta$ 
    is forced to be identical with the width of narrow [OIII]. The narrow H$\beta$ is referred as 
    $\rm{H\beta_{n}}$. The contribution, including the classical broad component and very broad component, 
    is referred as $\rm{H\beta_{b}}$. Figure 2 illustrates the modelling
    of the three observed spectra. The label of the spectrum in each panel from top to bottom  
    is the same as that in Figure 1. The observed profiles are represented by light solid lines,
    and the modelled profiles, by heavy solid lines. In each panel, the narrow core and broad base 
    of each line are shown by a short dashed line and a long dashed line, respectively.
    The residuals between the observed and the total fitted spectrum is displayed in the lower sub-panel
    underneath each spectrum.

    \section{Results and discussion}
    
    The three observations are compared with each other to investigate the variability of the profile 
    of broad component of [OIII]. The two recent observations span only about one year, and the observation
    in BG92 was performed round about ten years ago. Table 1 summarizes the line properties of the principal 
    emission lines at all epochs. In the following analysis and discussions in this paper, a constant flux of
    narrow core of [OIII] is assumed. Column 1 lists the indices of the observations. The indices are 
    same as the indices in Figure 1 and Figure 2. The flux ratios $\rm{H\beta_{n}/[OIII]_{n}}$ are listed
    in column 2, and the ratios $\rm{H\beta_{b}/[OIII]_{n}}$ in column 3. Column 4 lists 
    the normalized flux of broad [OIII] component. The velocity shifts of Gaussian peaks of broad [OIII]
    with respect to that of [OIII] narrow components are listed in column 5. A  negative velocity 
    corresponds to a blueshifted broad component relative to narrow component. The column 6 lists 
    the flux ratio of FeII emission to narrow [OIII], along with the upper and lower limits.
    The last column lists the calculated parameter RFe. RFe is defined as the flux ratio of the FeII
    complex to H$\beta$. We note that no attempt has been made to correct for the contamination of 
    the flux of narrow component of H$\beta$, because the flux of narrow H$\beta$ is expected to 
    be constant. This definition avoids the additional errors caused by the decomposition.
    In addition, the WWHI and WWHII averaged parameters denoted as the index WWH are calculated.
    The mean values of the interesting parameters are reported in the last row of Table 1, except 
    RFe, the ratio $\rm{H\beta_{b}/[OIII]_{n}}$ and $\rm{FeII/[OIII]_{n}}$.
    Note that the FWHM are excluded from Table 1, because we focus attention on flux variability 
    of the broad [OIII] component.

    \subsection{Variability of broad [OIII]}

    By comparing the results of BG92 and WWH, we find remarkable variations of profile of broad [OIII]
    during the period of ten years separating the observations. The normalized flux in broad [OIII]
    decreased from $1.90\pm0.22$ to $1.00\pm0.28$. In addition to the flux variability, the outflow(blueshift)
    $\Delta\upsilon_{r}([\rm{OIII}])$ increased from $-701.3\pm44.2\ \rm{km\ s^{-1}}$ to 
    $-959.7\pm104.6\ \rm{km\ s^{-1}}$. The change of velocity, WWH with respect to BG92,
    was -285.4\ $\rm{km\ s^{-1}}$ which is evidently larger than the errorbar estimated by the
    multi-component modelling.

    In Figure 3 we also show the residual profiles for all epochs as an additional test of line variability.
    The residual profiles are derived by removing the contributions of continuum, H$\beta$ 
    and narrow component of [OIII]. The residuals are also normalized by a constant flux of narrow [OIII]. 
    The vertical lines from bottom to top denote the centers of the modelled broad [OIII] components.

    The authenticity of the variations is verified by following two pieces of evidence. In the first
    instance, the normalized flux in broad [OIII] remained constant in the recent two epochs. 
    Once again, the coincident is elucidated by the superposition of the two residual spectra in Figure 3.
    According to this coincident, we are confident that the detected variations can not be caused 
    by the observational fluctuations. Secondly, the flux ratios $\rm{H\beta_{n}/[OIII]_{n}}$ are 
    constant approximately in the all epochs, which agrees with the prediction of generally
    accepted unified model of AGN. Therefore, we infer that the identified profile variations can 
    not be explained by the errors caused by the profile modelling.

    According to the variations of broad [OIII] during about 10 years separating the
    observations, we conclude that the physical length scale of broad [OIII] emitting does not
    exceed 10 light years, corresponding to $\sim$ 3pc. On the other hand, the broad [OIII] is 
    emitted from a lager region than the BLR($>1$ lyr), because it has not varied during the 1 year 
    separating the observations. Thus, the broad [OIII] emitting region is definitely situated at 
    outside the BLR. 

    The radius of the broad [OIII] emitting region can only be obtained 
    by indirect, model dependent arguments.  
    Assuming that the gravitational force of the central black hole dominates the 
    kinematics of the emitting region of broad [OIII], the broader profiles, with respect to the
    narrow profiles, indicate that the material emitting broad [OIII] should be located in 
    NLR inner regions. 
    The distance from central engine to the line region emitting broad [OIII] can be expressed as

    \begin{equation}
    R(\mbox{lt-days}) = 6.7\times10^{-6}\left(\frac{V_{\rm{FWHM}}}{10^{3}\ \rm{km\ s^{-1}}}\right)^{-2}
    \left(\frac{M_{\rm{BH}}}{M_{\odot}}\right)\cdot
    \end{equation}

    where $V_{\rm{FWHM}}$ is the FWHM of the emission profile of line emitting gas. The inferred
    black hole mass of I\,ZWI is $\log(M/M_{\odot})=7.26\pm0.11$. The value was estimated based upon single epoch optical
    spectroscopy(Vestergaard 2002). Adopting the averaged FWHM of the broad [OIII] component $\sim 1500\ \rm{km\ s^{-1}}$,
    the estimated distance is approximately about 100 lt-days.
  
    The inferred distance means that
    the broad [OIII] in I\,ZWI is most likely to be emitted in a transient emission line region(TLR) whose
    distance from central source is of order 100 lt-days. 
    In fact, it is a long time to realize that there appears to be some interconnection 
    between the classical BLR and NLR(Osterbrock \& Mathews 1986, Sulentic et al. 2000a). Following this concept,
    a few authors investigated the properties of the TLR. For instance, by examining the profile of high order
    Blamer series and [OIII], the TLR, with a scale size of order 1pc and an intermediate 
    FWHM($\sim1000\ \rm{km\ s^{-1}}$), was reported in X-ray selected AGN RE\,J1034+396(Mason et al. 1996).
    In the same object, by reproducing the observed spectra in terms of photoionization calculations, 
    Puchnarewicz et al.(1995) argued that the density of TLR($\sim10^{7.5}\ \rm{cm^{-3}}$) is much lower than
    that of BLR and higher than that of NLR.

    Up to now, in addition to the viral motions in gravitational potential of galaxy bugle(e.g. Nelson \&
    Whittle 1996), two other kinds of mechanism, radial flow from nucleus(for example, Crenshaw \& Kraemer et al. 2000)
    and radio jet expansion(e.g. Bicknell et al. 1998, Axon et al. 1998; Nelson et al. 2000),
    are suggested to explain the observed NLR. Both models can also give a reasonable origin of the 
    asymmetry of [OIII]. 
    For the first case, the observed peak+blue wing structure can be 
    interpreted as the emission from a radial outflowing component with a speed of a few hundreds $\rm{km\ s^{-1}}$ associated 
    with a disk component described by circular rotation(e.g. Kasier et al.(2000) in NGC\,4151; Veilleux et al.(2001) in 
    NGC\,2992). On the other hand, on the basis of the jet expansion assumption, 
    Nelson et al.(2000) modelled a set of long slit 2-dimensional spectra.
    This model reproduces the observed [OIII] profile in I\,ZW1, 
    if the far aside jet is obscured by galactic gas and the redshifted gas associated with near side jet lobe is optically 
    thick(see Fig.7 in Nelson et al.(2000)). Cecil et al.(2002) used this scenario 
    to analyze the deep spectra of NGC\,1068 taken at high spatial resolution.

     \subsection{Variation of FeII emission}

      In addition to the variations of the broad [OIII], the variability of the optical FeII emission
      is identified by examining Table 1(see column 6).
      We find a positive relation between the FeII emission and flux
      of H$\beta_{\rm{b}}$. According to the results of a great deal of AGN monitoring 
      studies, generally, the H$\beta$ flux closely relates to the continuum luminosity(e.g. Peterson et al. 2002,
      Kaspi et al. 2000, Kollatschny et al. 2000). Therefore, it is expected that 
      the emission of FeII increases with incident ionizing continuum in I\,ZWI. On the contrary,
      a marginally negative relation between RFe and H$\beta$(or continuum) flux is found in I\,ZWI.
      This means that in I\,ZWI the variations of the optical FeII blends are weaker in comparison to H$\beta$ line.

     \subsection{Low ionized NLR in I\,ZWI}

      The line ratios of the NLR in NLS1s are always interesting parameters. 
      The study of the NLR in NLS1 is not straightforward, however. Deblending the optical permitted lines 
      in NLS1 is difficult because no transition between the narrow and broad components is observed. 
      There is always large uncertainties in determining the fraction of H$\beta$ that is 
      emitted by the NLR. 
      In our analysis, both BG92 and WWH spectra indicate that the ratio $[\rm{OIII}]_{\rm{n}}$/H$\beta_{\rm{n}}$
      is approximate 1.7. 
      Such low value implies that the ionization stage of NLR in I\,ZWI is much lower than the generally
      adopted value($[\rm{OIII}]_{\rm{n}}$/H$\beta_{\rm{n}}\sim5-10$). 
   
      Is that an universal phenomenon in NLS1? As a matter of fact, V\'{e}ron-Cetty et
      al.(2004) found a very low ionized NLR in I\,ZWI from line system N3 in which most of oxygen might be 
      in the form of non-ionization, both because of the weakness and absence of 
      [OIII]$\lambda$5007, [SII]$\lambda\lambda$6716,6731 as well as [OII]$\lambda$7324 and because of the strength 
      of [CaII]$\lambda\lambda$7291,7324 lines. Furthermore,
      the low ionization stage in NLS1s was supported
      by the studies of Rodriguez-Ardila et al.(2000). They found that, on average in NLS1s, 50\% of flux of total
      H$\beta$ is due to emission from NLR and that the ratio $[\rm{OIII}]_{\rm{n}}$/H$\beta_{\rm{n}}$ varies
      from 1 to 5. The model analysis carried out by Contini et al(2003) demonstrated that such fraction of 
      the observed H$\beta$ line flux presents in the NLR flux in NLS1 galaxy Ark\,564. Additionally, Wang et al.(2004) 
      gave a ratio [OIII]/H$\beta_{\rm{n}}\sim$1.62 in SDSS\,J022119.84+005628.4 which is a recently identified NLS1 from Sloan 
      Digital Sky Survey. 
      All these results imply that normal broad line Seyfert 1s and NLS1s perhaps differ in ionization level of NLR.
      The NLS1s perhaps tend to commonly have a lower ionized NLR. 
      Although we do not know whether it is a truth, a vast of effort, of course, 
      will be made until the final result is achieved.

     \section{Conclusions}

     Two new spectra of NLS1 galaxy I\,ZW1 were recently taken by us and were compared with the spectrum 
     taken by BG92. The comparison allows us to make following conclusions:

      \begin{enumerate} 
      \item The variations of broad [OIII] are identified during the ten years separating
      the observations. The variations indicate that in I\,ZWI the length scale of emitting 
      of broad [OIII] does not exceed 3pc and is larger than 0.3pc. Assuming a Keplerian motion 
      in emission region, the material emitting broad [OIII] is possibly located in TLR, between BLR and NLR.

     \item A positive relation between the FeII emission and H$\beta$(or continuum) flux is 
     identified in the object I\,ZWI. Whereases, the parameter RFe marginally decreases 
     with ionizing continuum.

     \item In I\,ZWI, the inferred flux ratio of narrow component of [OIII] to narrow component of H$\beta$ is $\sim 1.7$
      rather than the generally adopted value($\sim5-10$). Such low value suggests that there is a low ionized NLR in I\,ZWI.

     \end{enumerate}

    \section{Acknowledgments}
  
    The authors thank Profs. Todd. A. Boroson and Richard. F. Green for providing the BG92's spectra
    and FeII template. We are also grateful to Dr. Y. F. Mao, D. W. Xu \& C. N. Hao for
    helpful discussions and suggestions. The special thanks go to the staff at Xinglong observatory for
    their instrumental and observing help. This work was supported by NSFC under grant 19973014.



   \clearpage 
   \begin{figure}
    \includegraphics[width=10cm]{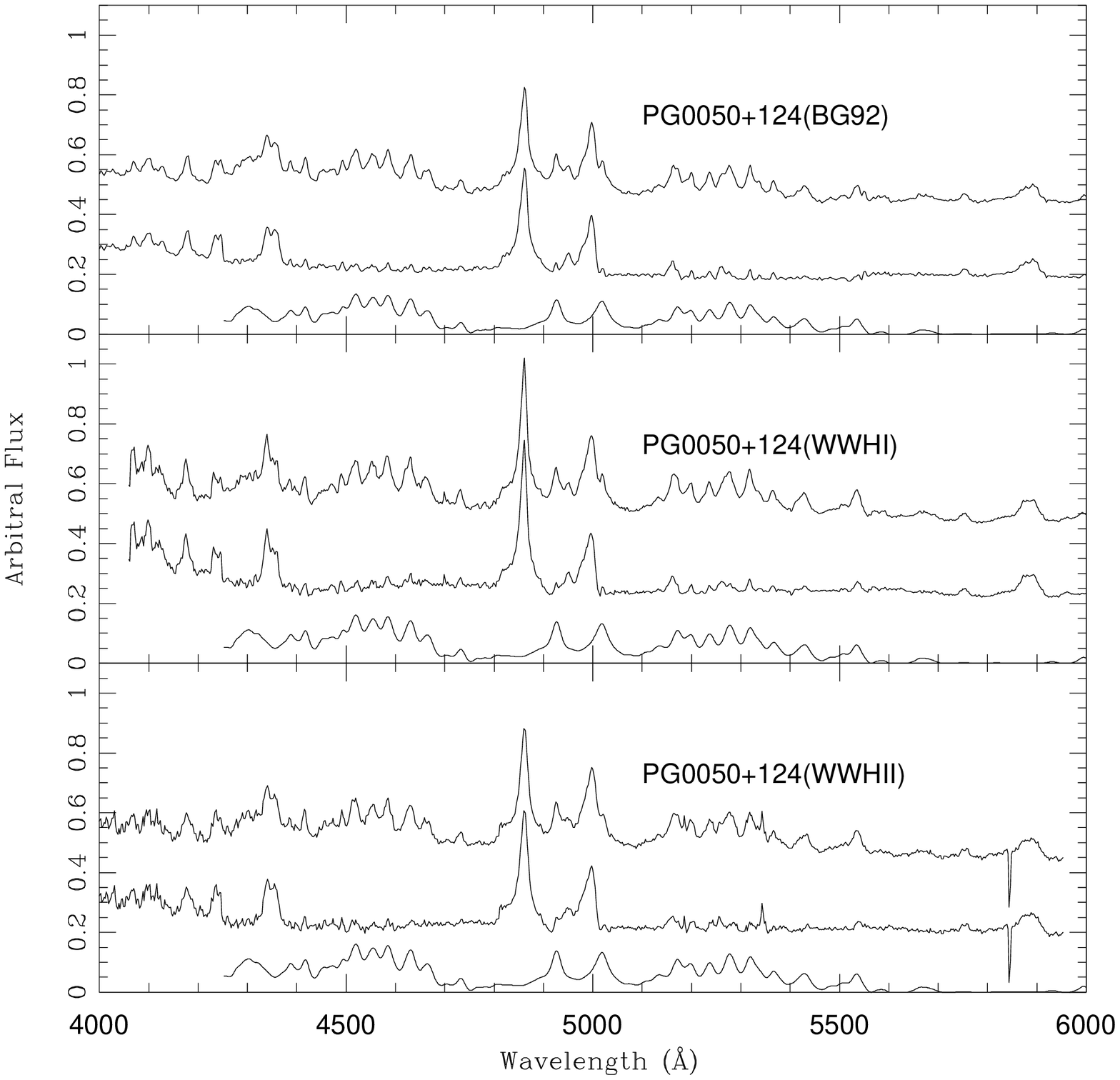}
    \caption{An illustration of the FeII subtraction for the three independent observations. From top to
     bottom panel, the spectra were taken by BG92, on Nov 16, 2001(hereafter denoted as WWHI) and on
     Dec 03, 2002(WWHII), respectively. In each panel, the top curve shows the observed spectrum which
     is shifted upward arbitrarily for visibility.
     The FeII-subtracted spectrum is shown by the middle curve.
     The bottom curve shows the best-fitted FeII template.
     }
    \end{figure}

    \begin{figure}
    \includegraphics[width=10cm]{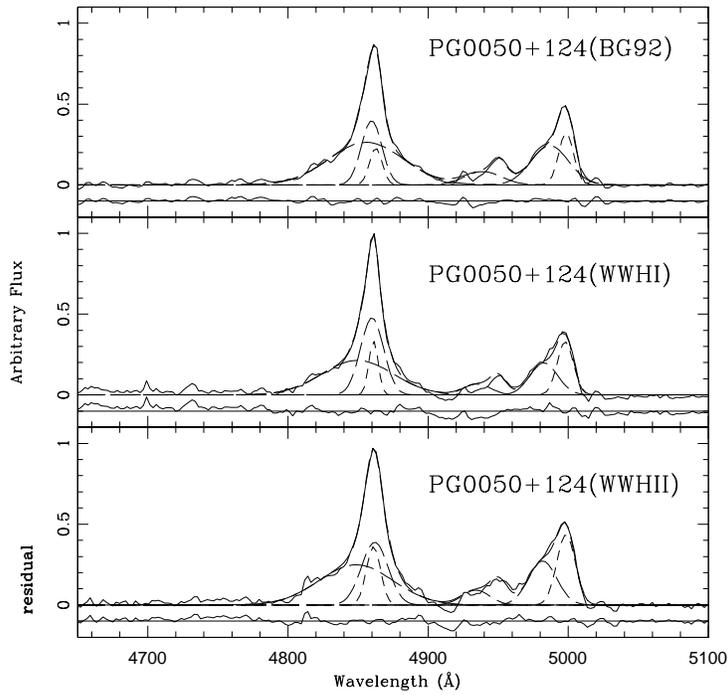}
    \caption{Profile modelling for three observations. The label of each spectrum from
      top to bottom is the same as in Figure 1. The light and heavy solid line represent the observed and modelled 
      profile, respectively. The narrow core of each line is displayed by short dashed line,
      the broad base by long dashed line. The residuals are shown by the lower sub-panel underneath each spectrum.}
    \end{figure}

   \begin{figure}
    \centering
    \includegraphics[width=10cm]{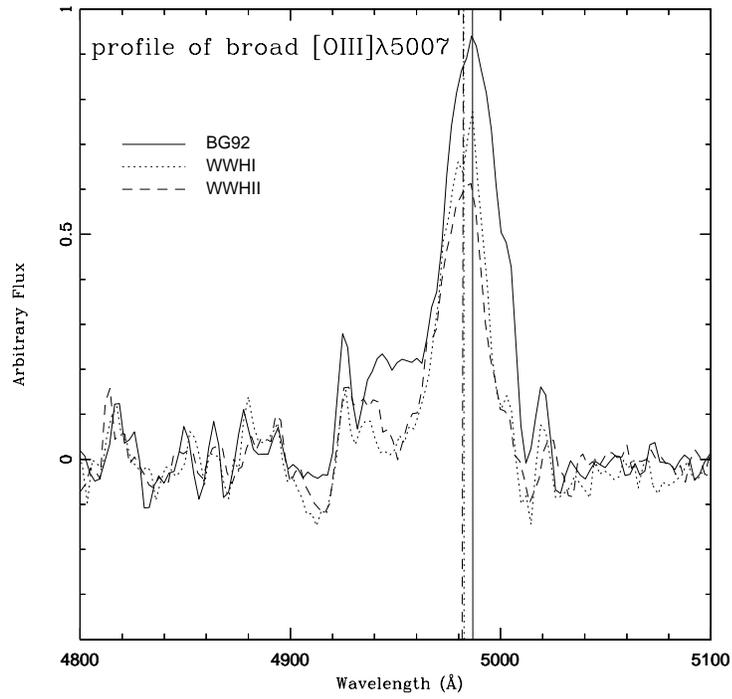}
    \caption{Comparison of residual profiles of broad [OIII] components. The positions of 
     center of modelled broad [OIII] are shown by lines from bottom to top.
     }
    \end{figure}

   \clearpage

   \begin{table}
    \caption{The Properties of Emission Lines of I\,ZWI}
      \label{1}
    $$
    \begin{array}{cccccccc}
    \hline
    \hline
    \noalign{\smallskip}
    \rm{Indices^{a}} & \rm{H}\beta_{\rm{n}}/[OIII]_{\rm{n}}
    &\rm{H\beta_{b}/[OIII]_{n}}  &  \rm{[OIII]_{b}/[OIII]_{n}}
    & \rm{\Delta\upsilon_{r}([OIII]})^{\rm{b}} & \rm{FeII/[OIII]_{n}} & \rm{RFe}\\
    (1) & (2) & (3) & (4) & (5) & (6) & (7) \\
    \noalign{\smallskip}
    \hline
    \rm{BG92} & 0.60\pm0.32 &  5.79\pm0.67 & 1.90\pm0.22 & -701.3\pm44.2 & 12.66^{+1.27}_{-1.27} & 1.98\\
    \rm{WWHI} & 0.59\pm0.87 &  4.96\pm1.04 & 1.04\pm0.27 & -919.1\pm95.6 & 10.96^{+0.91}_{-1.83} & 1.98\\
    \rm{WWHII}& 0.64\pm0.83 &  3.56\pm0.83 & 0.95\pm0.29 & -1000.2\pm112.8 & 8.96^{+0.74}_{-1.49} & 2.15\\
    \rm{WWH}  & 0.62\pm0.85 &  \dotfill    & 1.00\pm0.28 & -959.7\pm104.6 & ........ & \dotfill\\
    \noalign{\smallskip}
    \noalign{\smallskip}
    \hline

    \end{array}
    $$
    \begin{list}{}{}
    \item[$^{\rm{a}}$] The indices are identical with that in Figures 1 and 2.
    \item[$^{\rm{b}}$] The velocity shifts, in units of $\rm{km\ s^{-1}}$, of peaks of broad components
     with respect
     to that of narrow components, where a negative velocity corresponds to a blueshifted, broad component.
    \end{list}
   
   \end{table}


\begin{thebibliography}{00}





  \bibitem{axon} Axon, D. J., Marconi, A., Capetti, A., Macchetto, F. D., Schreier, E., \& Robinson, A., 1998, ApJ, 496, 75
  \bibitem{bicknell} Bicknell, G., Dopita, M. A., Tsvetanov, Z. I., \& Sutherland, R. S., 1998, ApJ, 495, 680
  \bibitem{boroson} Boroson, T. A., \& Green, R. F, 1992, ApJS, 80, 109
  \bibitem{busko} Busko, I. C., \& Steiner, J. E., MNRAS, 232, 525
  \bibitem{cecil} Cecil, G., Dopita, M., Groves, B., Wilson, A., Ferruit, P., P\'{e}contal, E., \& Binette, L.,
                2002, ApJ, 2002, 568, 627
  \bibitem{condon} Condon, J. J., Hutchings, J. B., \& Gower, A. C., 1985, AJ, 90, 1642
  \bibitem{crenshaw} Crenshaw, D. m., \& Kraemer, S. B., 2000, ApJ, 532, L101
  \bibitem{heckman} Heckman, T. M., Miley, G. K., van Breugel, W. J. M., \& Butcher, H. R.,
           1981, ApJ, 247, 403
  \bibitem{holt} Holt, J., Tadhunter, C. N., \& Morganti, R., 2003, MNRAS, 342, 227
  \bibitem{geupe} Grupe, D., Beuermann, K., Mannheim, K., \& Thomas, H. -C., 1999, A\&A, 350, 805
  \bibitem{kaspi} Kaspi, S., Smith, P. S., Netzer, H., Maoz, D., Jannuzi, B. T., \& Giveon, U., 2000, ApJ, 533, 631
  \bibitem{kaiser} Kaiser, M. E., Bradley, L. D., Hutchings, J. B., et al. 2000, ApJ, 528, 260
  \bibitem{kollatschny} Kollatschny, W., Bischoff, K., \& Dietrich, M., 2000, A\&A, 361, 901
  \bibitem{kriss} Kriss, G., 1999, ADASS, 3, 437
  \bibitem{marziani} Marziani, P., Sulentic, J. W., Zamanov, R., \& Calvani, M., 2003, Mem. S.A. It. Vol. 74, 490
  \bibitem{mason} Mason, K. O., Puchnarewicz, E. M., \& Jones, L. R., MNRAS, 283, 26
  \bibitem{massey} Massey, P., Strobel, K., Barnes, J. V., \& Anderson, E., 1988, ApJ, 328, 315
  \bibitem{nelson} Nelson, C. H., Weistrop, D., Hutchings, J. B., Crenshaw, D. M., Gull, T. R., Kaiser, S. B., \& Lindler, D., 2000, ApJ, 531, 257
  \bibitem{nelson} Nelson, C. H., \& Whittle, M., 1996, ApJ, 217, 415
  \bibitem{oke} Oke, J. B., \& Lauer, T. R., 1979, ApJ, 230, 360
  \bibitem{osterbrock} Osterbrock, D. E., \& Mathews, W. G., ARA\&A, 24, 171
  \bibitem{peterson} Peterson, B. M., et al., 2002, ApJ, 581, 197
  \bibitem{puchnarewicz} Puchnarewicz, E. M., Mason, K. O., Siemiginowska, A., \& Pounds, K. A., 1995,
          MNRAS, 276, 20
  \bibitem{rodriguez-ardila} Rodriguez-Ardila, A., Binette, L., Pastoriza, M. G., \& Donzelli, C. J.,
           2000, ApJ, 538, 581
  \bibitem{sergeev} Sergeev, S. G., Pronik, V. I., Malkov, Y. F., \& Chuvaev, K. K., 1997, A\&A, 320, 405
  \bibitem{storey} Storey, P. J., \& Zeippen, G. J., 2000, MNRAS, 312, 813
  \bibitem{sulentic} Sulentic, J. W., Marziani, P., \& Dultzin-Hacyan, D., 2000a, ARAA, 38, 521
  \bibitem{sulentic} Sulentic, J. W., Marziani, P., Zwitter, T., Dultzin-Hacyan, D., \& Calvani, M., 2000b,
           ApJ, 545, 15
  \bibitem{tadhunter} Tadhunter, C., Wills, K., Morganti, R., Oosterloo, T., \& Dickson, R., 2001, MNRAS, 327, 227
  \bibitem{veilleux} Veilleux, S., Shopbell, P. L., \& Miller, S. T., 2001, AJ, 121, 198 
  \bibitem{veron-cetty} V\'{e}ron-Cetty, M. -P., Joly, M., \& V\'{e}ron, P., 2004, A\&A, 417, 515
  \bibitem{veron-cetty} V\'{e}ron-Cetty, M, -P., V\'{e}ron, P., \& Goncalves, A. C., 2001, A\&A, 327, 730
  \bibitem{vestergaard} Vestergaard, M., 2002, ApJ, 571, 733
  \bibitem{xu} Xu, D. W., Komossa, S., Wei, J, Y., Qian, Y., \& Zheng, X. Z., ApJ, 2003, 590, 73
  \bibitem{wang} Wang, J., Wei, J. Y., \& He, X. T., 2004, ChJAA, 5, 415
  \bibitem{zamanov} Zamanov, R., Marziani, P., Sulentic, J. W., Calvani, M., Dultzin-Hacyan, D.,
          Bachev, R.,  2002, ApJ, 576, 9
  \bibitem{zwicky} Zwicky, F., 1971, Catalogue of selected compact galaxies and of post-eruptive galaxies, CIT, Pesadena


\end{thebibliography}
\end{document}